# Site-selective conductance of sidewall functionalized carbon nanotubes

Juan María García-Lastra[1], Kristian S. Thygesen[2], Mikkel Strange[2] and Ángel Rubio[1]

[1]*Nano-Bio Spectroscopy Group and European Theoretical Spectroscopy Facility (ETSF), Departamento de Fisica de Materiales, Unidad de Materiales Centro Mixto CSIC-UPV/EHU, Universidad del Pais Vasco, Avd. Tolosa 72, E-20018 Donostia, Spain*

[2]*Center for Atomic-scale Materials Design (CAMD), Department of Physics, Technical University of Denmark, DK-2800 Kgs. Lyngby, Denmark*

**Abstract**

We use DFT to study the effect of molecular adsorbates on the conductance of metallic carbon nanotubes. The five molecules considered ($NO_2$, $NH_2$, H, COOH, OH) lead to similar scattering of the electrons. The adsorption of a single molecule suppresses one of the two available channels of the CNT at low bias conductance. If more molecules are adsorbed on the same sublattice, the remaining open channel can be blocked or not, depending on the relative position of the adsorbates. If a simple geometric condition is fulfilled this channel is still open, even after adsorbing an arbitrary number of molecules.

Sidewall chemical functionalisation of carbon nanotubes (CNT) has been the subject of several studies (see reviews in 1-3) as it represents a very direct way to implement a number of technological applications including nanoelectronics devices and various types of sensors[3]. These applications all build on the idea that the electrical conductivity of the nanotube can be controlled externally by adding/removing different functionalised groups[4].

Much theoretical work has been devoted to the analysis of the impact of lattice symmetry and potential impurity/defect scattering in the I/V characteristics of CNTs[1, 3, 5-7], not only in single wall CNTs, but also in double-wall CNTs[8]. Those studies put in evidence that the number of available conducting channels in metallic nanotubes depends on the range of the scattering and the chirality of the tube. However, very little is known about the impact of bipartite-lattice symmetry of graphene when multiple scattering centers are present.

In the present work we show that the transport properties of functionalized CNTs in the phase-coherent regime show a peculiar, systematic dependence on the positions of the adsorbed molecules. Namely, for any number of molecules adsorbed on the *same* sublattice, the conductivity of the tube remains close to $1G_0$ if the relative positions of the adorbates satisfy a simple geometric condition, whereas the conductance is strongly suppressed if one or more molecules fail to satisfy the condition. As a second result, we show that the change in conductivity of a metallic CNT due to chemisorption of a molecule at an on top site of the nanotube sidewall is largely independent of the molecular specie. This universal behavior is verified for H, COOH, OH, $NH_2$, and $NO_2$ adsorbed on (6,6) armchair, (9,0) zigzag, and (8,2) chiral CNTs.

Recall that a graphene sheet contains two inequivalent types of sites, A and B, constituting the two sublattices of graphene. The $p_z$ orbitals of the carbon atoms form the π- and π*- valence and conduction bands which intersect at the K and K' points of the Brillouin zone (BZ). When a graphene sheet is rolled up to form a CNT, the band structure (neglecting curvature effects) is obtained by restricting the band structure of graphene to discrete lines in the BZ. Metallic CNTs thus have two available channels for transporting electrons (in both directions along the tube axis) at energies close to the Fermi level formed by the Bloch states in the vicinity of the K and K' points. We shall refer to these states as forward- and backward-moving states and denote them by $|K_+>$, $|K'_+>$ and $|K_->$, $|K'_->$, respectively.

Within a Tight-Binding (TB) model we can now address how the nanotube electronic response is perturbed by the presence of one molecule adsorbed on the CNT sidewall. To this end we consider the TB model of the CNT with a single level of energy $\varepsilon_0$ coupled to one of the $p_z$ orbitals of the CNT via the hopping parameter $t$. We can take $\varepsilon_0$ to be negative or positive corresponding to a model for the HOMO or LUMO of molecule. Within this model, which is a particular case of Fano resonance[9], we can calculate the elastic transmission function of the CNT. We have done this using the non-equilibrium Green's function (NEGF) formalism[10] as described in Refs[11, 12].

The calculated transmission function of the model[13] exhibits a dip in one of the two transport channels. The position of the dip is directly proportional to $\varepsilon_0$. For weak coupling, i.e. small values of $t$, the dip becomes more narrow and shallow and is located closer to $\varepsilon_0$, while for strong coupling the dip is wider with a depth approaching 1 (which means that one of the two transmission channels has been completely suppressed). In the strong coupling case the dip is located close to the onsite energy of the CNT sites which coincide with the Fermi level.

We can understand this result and predict what happens when two molecules are chemisisorbed, by rather simple means. Let us consider the subspace of states spanned by the four degenerate CNT Bloch states at Fermi level and the impurity state(s). The Hamiltonian matrix within this subspace is shown below in the case of a single impurity adsorbed at $\vec{R}=0$, and two impurities adsorbed at $\vec{R}=(0,0)$ (an A site) and other at the general site $\vec{R} = n \cdot \vec{a}_1 + m \cdot \vec{a}_2$, respectively:

$$i) \begin{pmatrix} \Delta & 0 & 0 & 0 & t \\ 0 & \Delta & 0 & 0 & t \\ 0 & 0 & \Delta & 0 & t \\ 0 & 0 & 0 & \Delta & t \\ t & t & t & t & \varepsilon_0 \end{pmatrix}; \quad ii) \begin{pmatrix} \Delta & 0 & 0 & 0 & t & t \cdot f_{K_+} \\ 0 & \Delta & 0 & 0 & t & t \cdot f_{K_-} \\ 0 & 0 & \Delta & 0 & t & t \cdot f_{K'_+} \\ 0 & 0 & 0 & \Delta & t & t \cdot f_{K'_-} \\ t & t & t & t & \varepsilon_0 & 0 \\ t \cdot f^*_{K_+} & t \cdot f^*_{K_-} & t \cdot f^*_{K'_+} & t \cdot f^*_{K'_-} & 0 & \varepsilon_0 \end{pmatrix} \quad (1)$$

where $f_i$ is the phase factor of the Bloch state at the position of the second impurity ($f_i$ values are detailed in the section 2 of Ref[13]). The parameter $2\Delta$ represents the gap induced by the curvature in zigzag and chiral CNTs ($\Delta=0$ for armchair CNTs) The eigensolutions of matrix $i)$ are shown in table 1. The eigenvectors 3, 4, and 5 represent non-propagating states (they have equal weight on left- and right-moving Bloch states), while the eigenvectors 1 and 2 represent states propagating in the left and right directions, respectively. It is important to notice that the eigenvectors having zero weight on the impurity, in particular states 1 and 2, are not only solutions in the restricted subspace but are in fact exact solutions to the full problem as they do not couple to the remaining Bloch states of the CNT. Consequently, state 1 (2) constitutes a channel for transporting electrons unhindered through the tube in the left (right) direction even in the presence of the impurity.

The solutions of the secular problem *ii)* depend strongly on the position of the second impurity. We can distinguish three cases:

**1)** When the second impurity is at an A site such that the relative distance R fulfils the condition:

$$R = n \cdot \vec{a}_1 + m \cdot \vec{a}_2, n - m = 3p, \text{ p being an integer} \qquad (2)$$

all the $f_i$ parameters are 1 (for every kind of metallic CNTs) and the corresponding eigenvectors of *ii)* have the same form as those of *i)*. Thus, in this case there are again two fully transmitting eigenstates (moving in the positive and negative directions of the tube, respectively) which can carry electrons unhindered through the CNT. The extra eigenvector as compared to *i)* corresponds to the antibonding combination of the two impurity levels, without any weight from the CNT. It is instructive to notice that (2) is the same condition required by a CNT to be metallic[14, 15] (neglecting curvature/hybridization effects).

**2)** If the second impurity is placed at an A site, but does not satisfy the condition (2) all the eigenvectors are non-propagating (for every kind of metallic CNTs), and the transmission is completely suppressed.

**3)** If the second impurity is placed at a B site the form of the eigenvectors depend strongly on the actual adsorption site. While there will be eigenstates with a non-zero momentum along the tube axis, none of these will be fully transmitting, i.e. have weight solely on either forward- or backward moving Bloch states. Consequently the conductance is expected to fluctuate more or less randomly as the second adsorption site is varied over the B sublattice.

*More generally it can be shown that the propagating eigenstate* $|\psi\rangle = -\frac{1}{\sqrt{2}}|K'_+\rangle + \frac{1}{\sqrt{2}}|K_+\rangle$ *(eigenstate 2 of table 1) vanishes simultaneously at all the A sites fulfilling the condition* (2) *and thus enables scattering-free transport though the*

*CNT in the presence of an arbitrary number of impurities adsorbed at these special sites of the A sublattice.*

We mention that similar models, including the **k.p** equation and the Born approximation for point like scattering potentials, have previously been used to study impurity scattering in CNT[6].

In order to address the validity of the conclusions drawn from the simple TB model for scattering under more realistic conditions, we have performed first principles DFT+NEGF transport calculations following the approach described in Refs.[11, 12] for a number of small molecules (H, COOH, OH, $NH_2$ and $NO_2$) chemisorbed on the (6,6) armchair, (9,0) zig-zag, and (8,2) chiral CNTs[13]. We have chosen these CNTs due to their small radii which make them more reactive towards binding molecules at the side-wall as compared to large radius CNTs[16] (curvature induces small $sp^3$-like hybridization).

All molecules bind covalently to an on-top site of the CNT wall. The calculated distances between the adsorbates and the CNT, $d_i$, are (in Å) $d_H$=1.18, $d_{COOH}$=1.57, $d_{OH}$=1.45, $d_{NO_2}$=1.64 and $d_{NH_2}$=1.49. We have also calculated the distances and transmission curves for CO and $CO_2$. These molecules are physisorbed by the CNT at $d_{CO}$=2.55 and $d_{CO_2}$=2.52, and we have found that they do not influence the transport properties of the CNT close to the Fermi level[13]. For this reason we focus on the chemisorbed species in the following.

In Fig. 1 we show the effect of a single chemisorbed molecule on the transmission function for the different CNTs. For all molecules the effect is to reduce the transmission function from $2G_0$ to $1G_0$ in the vicinity of the Fermi level, in agreement with the TB results for the strong coupling case (see Fig. 2 in the supplementary material [13]). The same effect was observed by Lee et al. for the phenyl molecule adsorbed on top of a carbon atom in a (5,5) CNT[16]. This happens because one of the two channels which are available in the pure CNT around the Fermi level is almost completely suppressed by scattering off the adsorbate while the other remains unaffected. This is clearly seen by decomposing the transmission function into the non-mixing eigenchannels[17]. The contributions from individual eigenchannels to the total transmission are shown in the Fig. 1 b). In Fig. 2 (a) and (b) we show the contour surfaces of the fully transmitting and blocked eigenchannels at the Fermi energy. Notice that the fully transmitting channel vanishes at A sites fulfilling condition (2) as predicted by the TB model.

It is interesting to notice that, despite the fact that the molecules bind to the CNT via an H, C, O, and N atom, respectively, the effect of the adsorbates on the transmission is very similar, the only difference being a slight shift in the position of the dip. Clearly, these results are in line with the predictions of the TB model the strong coupling regime[13].

We next consider what happens when two molecules are adsorbed at different sites of the CNT. We adsorb the first molecule at an A site, and vary the adsorption site for the second molecule. We distinguish between the cases where both molecules are located on A sites and cases where the second molecule is adsorbed at a B site. We have performed calculations for various adsorption sites for all molecular species as well as for combinations of those molecules. As in the case of a single molecule, the shape of the transmission function is insensitive to the adsorbed species. Again this indicates that the transport properties of metallic CNTs are affected in a similar way by different chemisorbed molecules. Having observed this, we focus to the simplest case of adsorbed H in what follows. The results for H adsorbed at different A-A configurations on the armchair, zig-zag, and chiral CNTs are shown in Fig. 3. It is easy to distinguish two different types of transmission curves: For the cases fulfilling Eq. (2) the transmission at the Fermi level is close to $1$[18]. For the cases not fulfilling Eq. (2) the transmission at the Fermi level almost completely suppressed (less than $0.1G_0$)[19]. In contrast there is no trend in the transmission functions of the A-B configurations[13], and the transmission values at the Fermi level fluctuate randomly between 0 and $2G_0$. Similar fluctuations were observed by Lee et al., who considered only A-B configurations of phenyl pairs[16].

We have made several tests to study the robustness and universality of the site-selective conductivity. When two different molecular species are adsorbed, the transmission dips become slightly wider in all cases. In some of the cases where Eq. (2) is not fulfilled, the transmission is not always fully blocked, but has a minimum of $0.25G_0$ rather than $0.1G_0$. This is due to the different position of the HOMO levels of the two molecules. We also analysed the effect of temperature by making random displacements of the atoms corresponding to 300 K and found only minor changes in the transmission functions. However, replacing the adsorbates by vacancies produces more long ranged perturbations and destroys the effect described here. Although the present calculations are based on an independent particle approximation, electron-electron interactions are not expected to affect the general trends discussed here, as the follow mainly from the symmetry of the lattice and the wavefunctions (see Ref[20] and references therein for a discussion of many-body effects in quantum transport in nanostructures).

There are two main conclusions of this Letter. First, when a single molecule is chemisorbed on the sidewall of a metallic CNT, one of its two transmission channels close to the Fermi level is blocked while the other remains unaffected. We have found that this behavior is quite universal in the sense that the depth and position of the transmission minimum depends only weakly on the molecular species. Secondly, it is possible to adsorb an arbitrary number of molecules on the CNT sidewall and still conserve a fully open transport channel. To obtain this, all molecules must be adsorbed on the same sublattice and their relative positions should fulfill Eq. (2). On the other hand, if a single molecule is adsorbed outside this array of special sites, both channels are almost completely blocked and the conductance drops sharply, independently of tube chirality.

[18] In comparison with the case of one single molecule adsorbed (Fig. 1) the width of the dip is broadened, due to the fact that the suppressed channel is scattered by two molecules. However, as in Fig. 1, one channel remains completely open.

[19] We have checked these properties when the two molecules are in different unit cells and also for the case with three molecules in the same sublattice. Again one channel remains open if condition 2 is satisfied and both channels are blocked if it is not.

[20] K. S. Thygesen and A. Rubio, Physical Review B **77** (2008).

**Table 1.** Eigenvalues and (unnormalized) eigenvectors obtained after solving eq. (1) *i)*. The eigenvectors are expressed as linear combinations of the four degenerate valence band eigenvectors at the grapheme Brillouin Zone and the impurity level, $|\varepsilon_0\rangle$.

$$E^{\pm} = \frac{\varepsilon_0 - \Delta \pm \sqrt{\varepsilon_0^2 + \Delta^2 + 16t^2 + 2\Delta\varepsilon_0}}{2}.$$

|    | Eigenvalue | $|K_+\rangle$ | $|K'_-\rangle$ | $|K'_+\rangle$ | $|K_-\rangle$ | $|\varepsilon_0\rangle$ |
|----|------------|---------------|----------------|----------------|---------------|-------------------------|
| 1) | $-\Delta$  | 0             | 1              | 0              | -1            | 0                       |
| 2) | $-\Delta$  | -1            | 0              | 1              | 0             | 0                       |
| 3) | $-\Delta$  | -2            | 1              | 0              | 1             | 0                       |
| 4) | $E^+$      | $t/(E^+-\Delta)$ | $t/(E^+-\Delta)$ | $t/(E^+-\Delta)$ | $t/(E^+-\Delta)$ | 1 |
| 5) | $E^-$      | $t/(E^--\Delta)$ | $t/(E^--\Delta)$ | $t/(E^--\Delta)$ | $t/(E^--\Delta)$ | 1 |

**Figure captions**

**Figure 1.** Up-Left: Transmission curves for a (6,6) carbon SWNT with a single molecule chemisorbed on-top of a C atom: COOH (black), H (red), $NH_2$ (green), $NO_2$ (blue) and OH (orange). The transmission curve for the pure nanotube is also shown (black dashed line). Down-Left: The same for a (9,0) CNT. Down-Right: The same for a (8,2) CNT. Up-Right: Transmission of the different eigenchannels for a (6,6) CNT with a single COOH molecule adsorbed. Notice that channel 1 (solid line) is completely unaffected by the impurity, while channel 2 (dashed line) is completely suppressed in the vicinity of the Fermi level.

**Figure 2.** Contour surfaces of the two eigenchannels at the Fermi level of a (6,6) CNT with a single hydrogen atom adsorbed on-top of a C atom. The states correspond to channels 1 and 2 in the upper-left panel of Fig. 1. The position of the hydrogen atom is indicated by the blue arrow. The CNT is shown unrolled for clarity. (a) Fully transmitting eigenchannel, $T=1$. (b) Blocked eigenchannel, $T=0$. Notice that channel 1 has zero weight on all sites fulfilling Eq. (2).

**Figure 3.** Top: Transmission functions for a (6,6) carbon SWNT with two hydrogen atoms adsorbed at different A sites. The six inequivalent A-A configurations in one unit cell of the CNT are shown. The blue solid lines correspond to the cases where Eq. (2) is fulfilled. The red dashed lines correspond to the cases where Eq. (2) is not fulfilled. The transmission function for the pure CNT is also shown (black dashed line) Middle: The same for a (9,0) CNT. We show the 9 inequivalent A-A configurations in one unit cell of the CNT. Bottom: The same for a (8,2) CNT. We show 6 out of the 14 inequivalent A-A configurations within a CNT unit cell.

**Figure 1. Garcia-Lastra et al.**

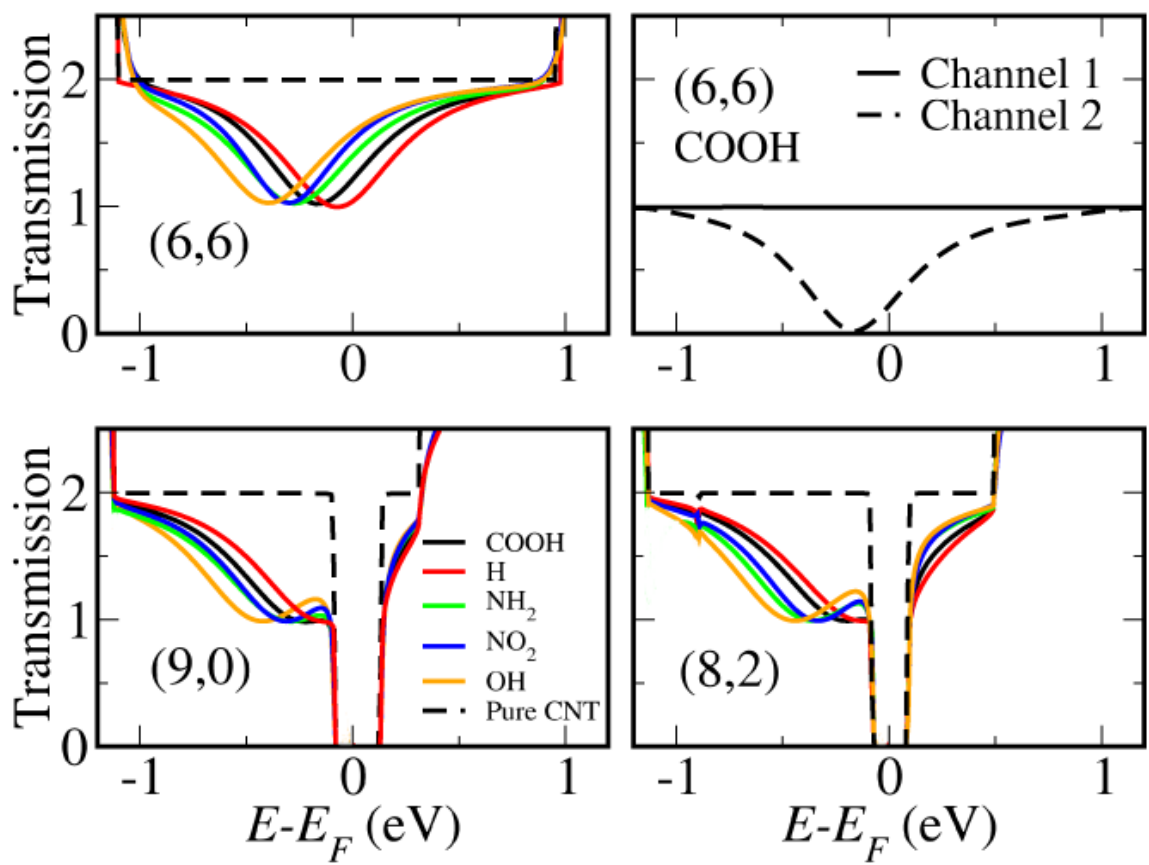

**Figure 2. Garcia-Lastra et al.**

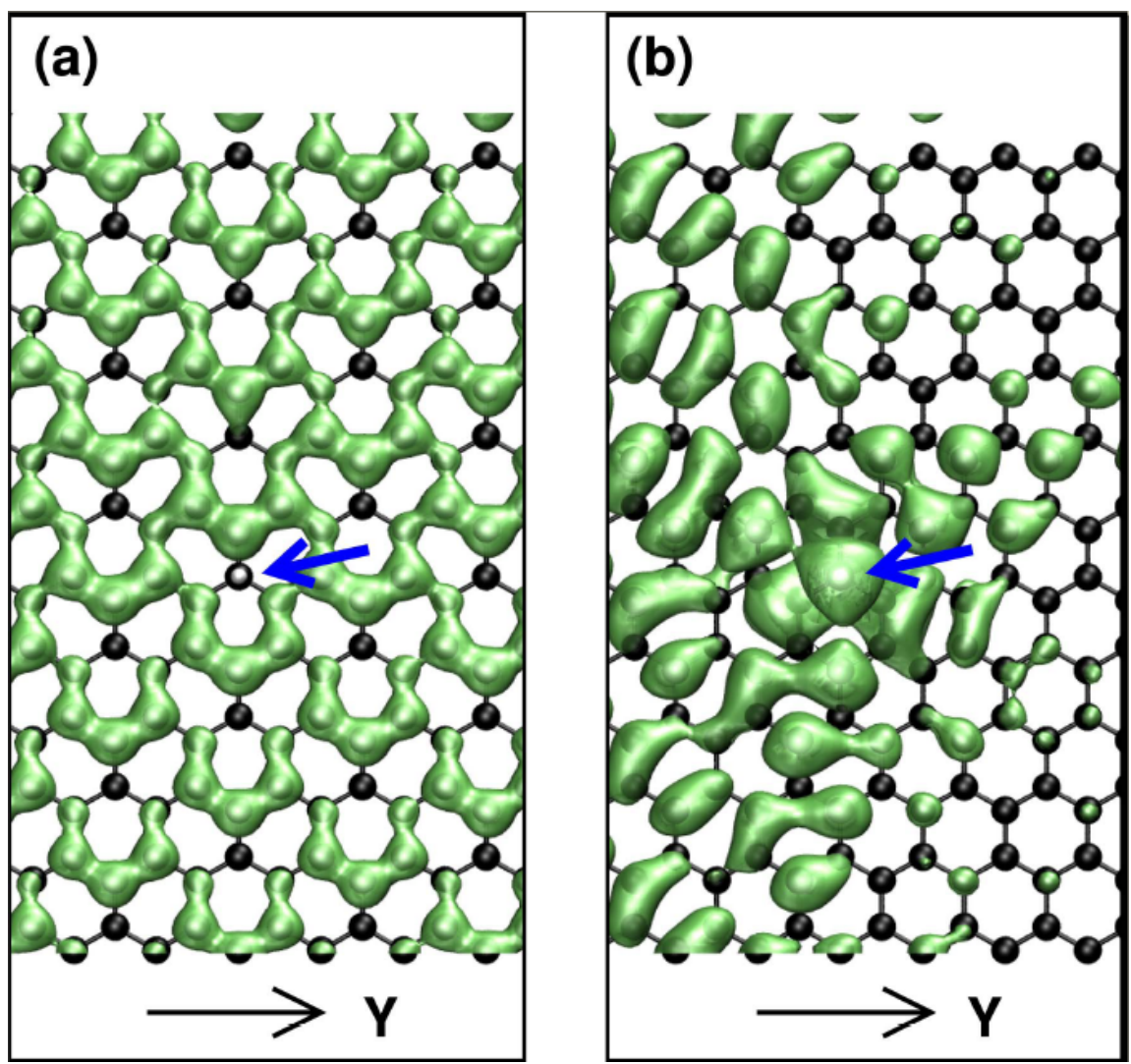

**Figure 3. Garcia-Lastra et al.**

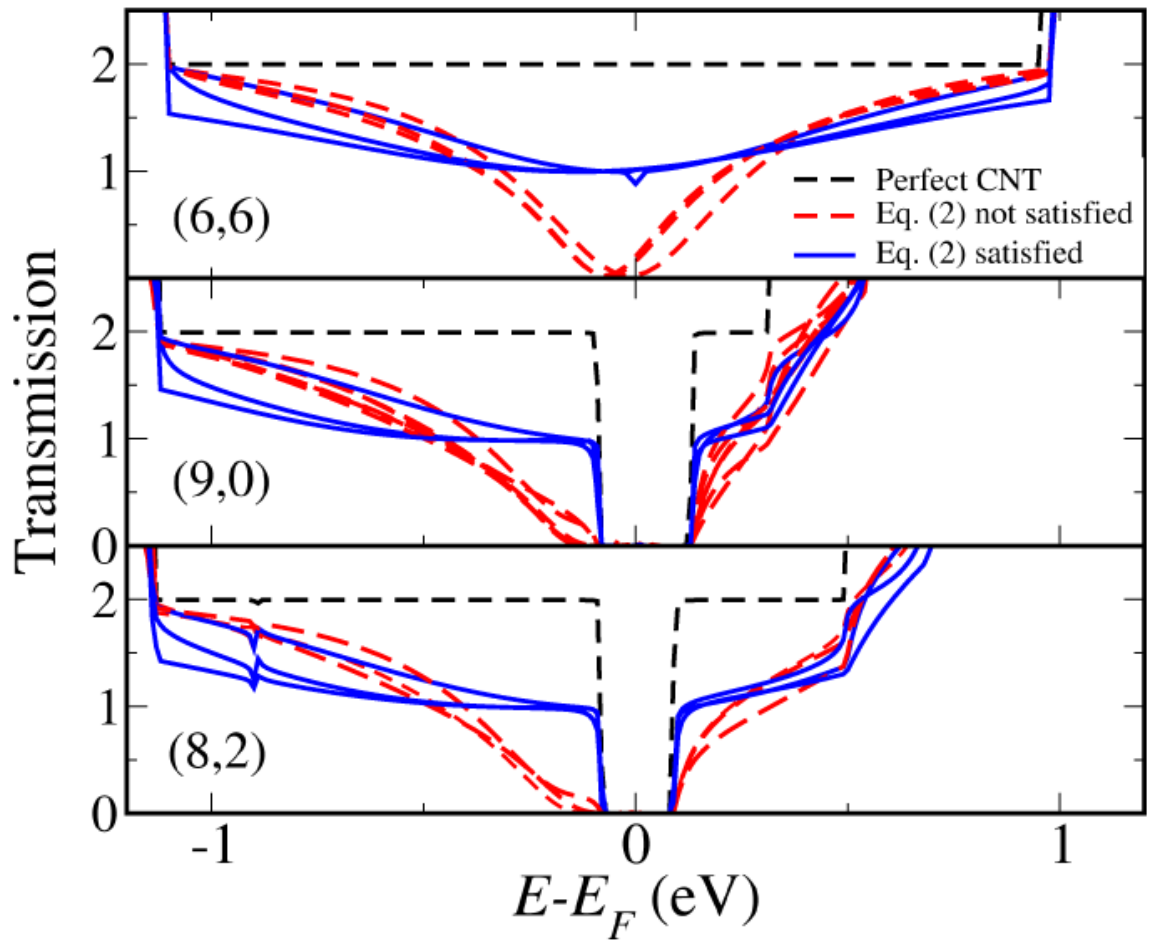

# Supplementary Material for

## "Site-selective conductance of sidewall functionalized carbon nanotubes"

by

Juan María García-Lastra, Kristian S. Thygesen, Mikkel Strange and Ángel Rubio

## 1. Calculation details

The geometry optimizations have been performed without any imposed constrain by means of Siesta[1] code. The supercells contain six unit cells for the (6,6) CNT (144 atoms without dopants), six for the (9,0) CNT (216 atoms without dopants) and three for the (8,2) CNT (168 atoms without dopants). Only Γ point was considered. For exchange and correlation we used the LDA functional of Perdew and Zunger[2] (PZ). A single-zeta polarised (SZP) basis set was used for all atomic species. All the calculations are non spin polarized. We have verified that the transport results are not affected by enlarging the basis set, enlarging the size of the supercell, or by including spin polarization

## 2. Structure of graphene and CNTs and its influence on $f$ factors of equation 1.

Figure 1 shows the basic features of the graphene and the CNTs. It is important to notice that the allowed lines in the Brillouin zone (the bands responsible of the conduction close to the Fermi Level) in the case of an armchair CNT connect a K point with a K' point (Fig 1b), while they connect a K (K') point with a K (K') point in the case of zigzag and chiral nanotubes (Fig 1c). For this reason $f_i$ parameters are different in the case of armchair CNTs and in the case of zigzag or chiral CNTs. $f_i$ parameters depends on the position of the second impurity as $f_{K_+} = e^{-i\frac{2\pi}{3}(n-m)}$, $f_{K_-} = e^{i\frac{2\pi}{3}(n-m)}$, $f_{K'_+} = e^{i\frac{2\pi}{3}(n-m+3m)-i\pi\cdot s}$ and $f_{K'_-} = e^{-i\frac{2\pi}{3}(n-m-3n)-i\pi\cdot s}$ for the armchair CNTs and $f_{K_+} = e^{-i\frac{2\pi}{3}(n-m+3m)+i\frac{\pi}{2}\cdot s}$, $f_{K_-} = e^{-i\frac{2\pi}{3}(n-m-3n)-i\frac{\pi}{2}\cdot s}$, $f_{K'_+} = e^{i\frac{2\pi}{3}(n-m-3n)+i\frac{\pi}{2}\cdot s}$ and $f_{K'_-} = e^{i\frac{2\pi}{3}(n-m+3m)-i\frac{\pi}{2}\cdot s}$ for the zigzag and chiral CNTs. In all the cases the parameter $s$ is 0 if the second impurity is at an A site and 1 if it is at a B site.

Apart from the references of the main text, relevant information about the structure and symmetry of CNTs (pure and doped) and its influence on the transport properties can be found in the literature[3-13]. The series of papers by Ando[3-6] are specially related with the symmetry aspects treated in the main text.

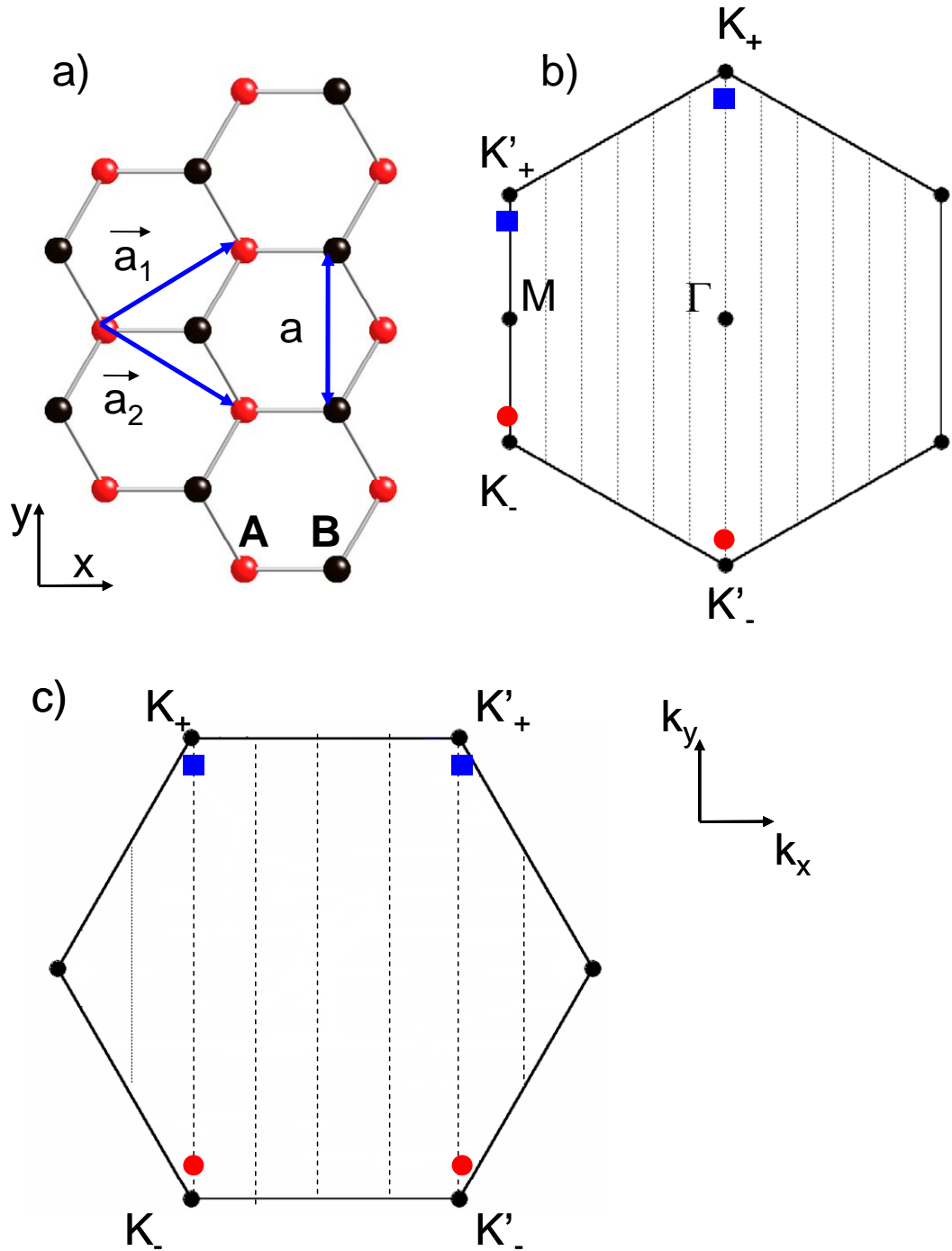

**Figure 1.** (a) Basis vectors of the hexagonal lattice of graphene, $\vec{a}_{1,2} = \frac{a}{2}(\sqrt{3}, \pm 1)$, with $a = \frac{2}{\sqrt{3}} d_{c-c}$. Atoms in A (B) sites are in red (black). The coordinates of the A and B sites in the unit cell at the origin are $(0,0)$ and $\frac{1}{3}(\vec{a}_1 + \vec{a}_2)$, respectively. (b) The first Brillouin zone of a (6,6) armchair carbon SWNT with the K and K' points at $K_+, K'_- = \frac{4\pi}{3a}(0, \pm 1)$. (c) The first Brillouin zone of a (9,0) zigzag carbon SWNT with the K and K' points at $K_+, K_- = \frac{2\pi}{3a}(-1, \pm\sqrt{3})$. The blue squares and red circles correspond to states moving in the positive and negative directions of the tube axis, respectively (see also (b)). The conductance takes place along y axes.

## 3. Tight-binding results

Figures 2 and 3 show the dependency of the transmission curves on the hopping parameter, $t$, and the relative position of the HOMO of the adsorbate with respect to the Fermi level of the CNT, $\varepsilon_0$, in the tight-binding model. As it is said in the main text, for weak coupling, the dip becomes more narrow and shallow and is located closer to $\varepsilon_0$, while for strong coupling the dip is wider with a depth approaching 1 and is located closer to the Fermi level of the CNT (Fig 2). The weak coupling regime corresponds with the case of physisorbed molecules (see Fig. 4). On the contrary, all the chemisorbed molecules studied in the present work belong to the strong coupling regime. On the other hand it could be seen in Fig. 3 that the position of the minimum of the dip is proportional to $\varepsilon_0$.

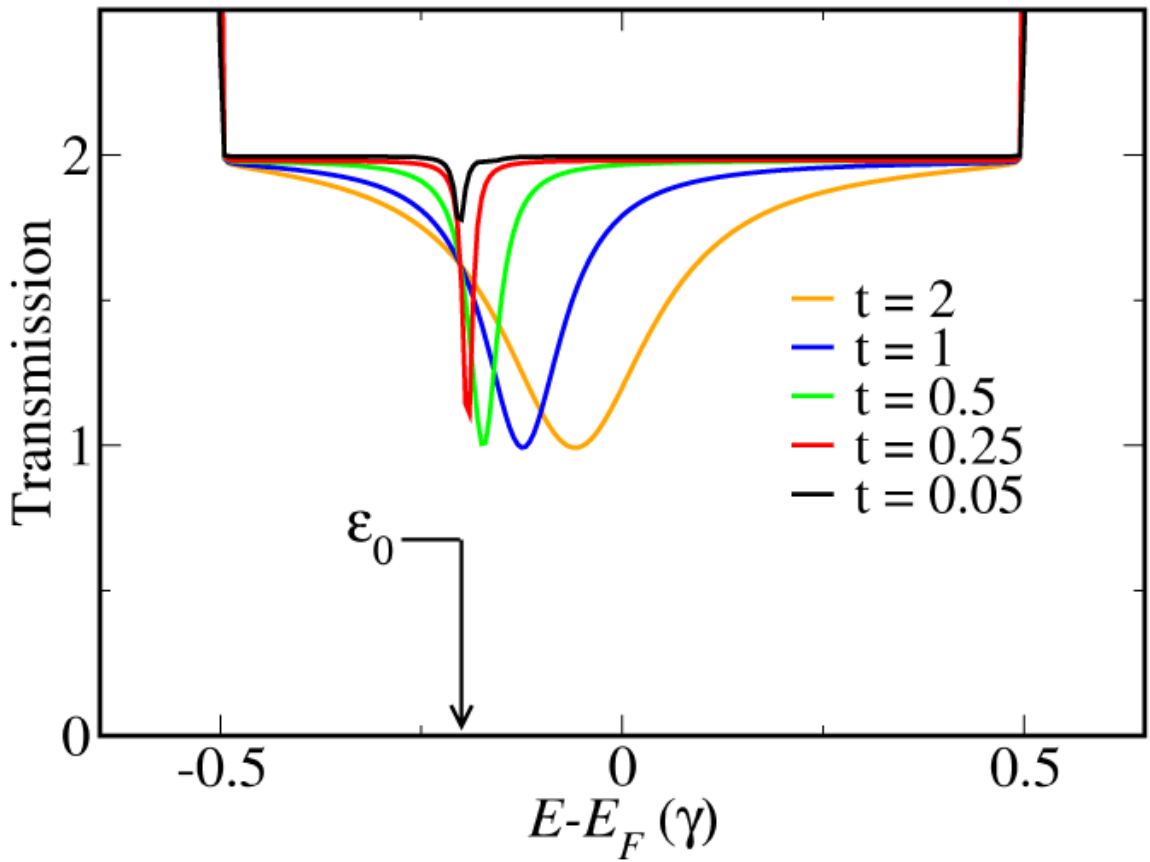

**Figure 2.** Transmission curves for a (6,6) CNT doped with a single adsorbate obtained through the tight-binding model for different values of the hopping energy parameter, t: t=0.05 (black), t=0.25 (red), t=0.5 (green), t=1 (blue), t=2 (orange) when the relative position of the HOMO of the adsorbate with respect to the Fermi Level of the CNT is $\varepsilon_0$ = -0.2 $\gamma$ (it is marked with an arrow). Energy units in $\gamma$ (the energy of the hopping integral between the $p_z$ orbitals of two adjacent carbon atoms in a CNT)

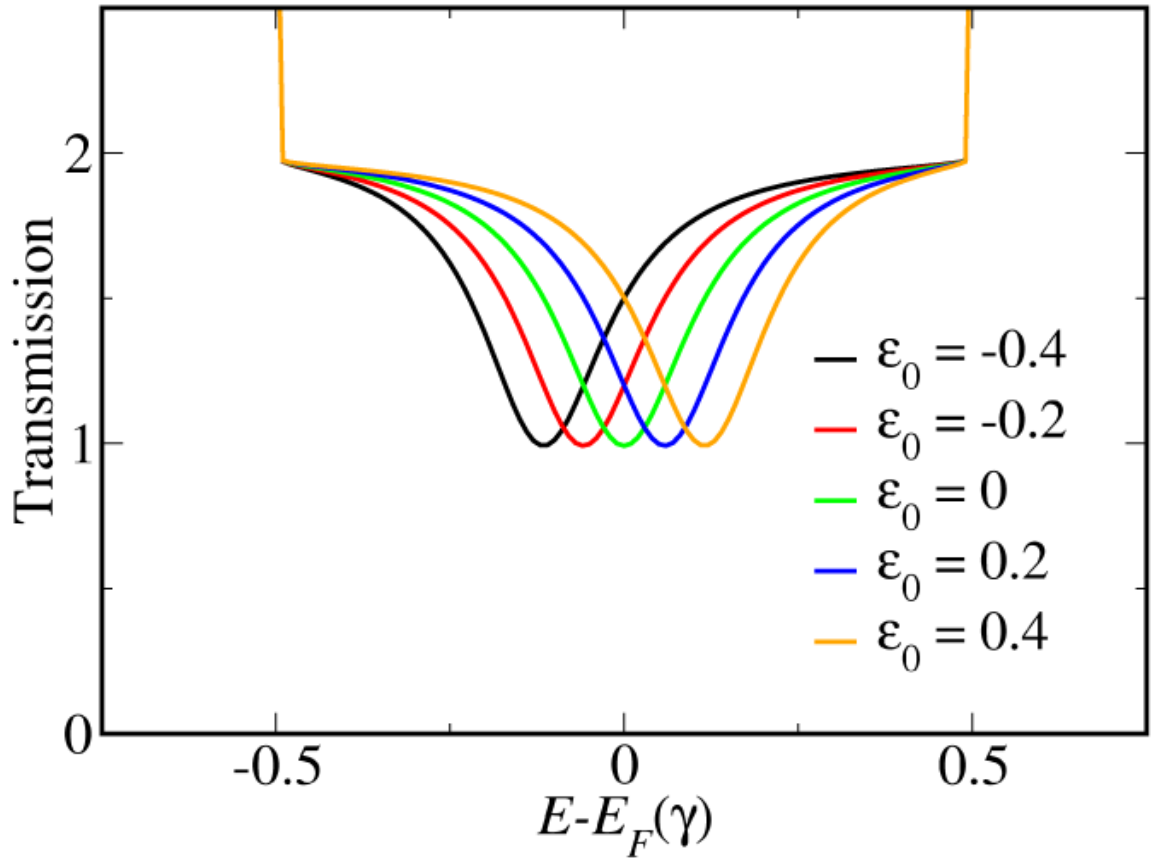

**Figure 3.** Transmission curves for a (6,6) CNT doped with a single adsorbate obtained through the tight-binding model for different values of the relative position of the HOMO of the adsorbate with respect to the Fermi Level of the CNT $\varepsilon_0$: $\varepsilon_0$=-0.4 (black), $\varepsilon_0$=-0.2 (red), $\varepsilon_0$=0 (green), $\varepsilon_0$=0.2 (blue) and $\varepsilon_0$=0.4 (orange), when the hopping energy parameter is t=2$\gamma$. Energy units in $\gamma$ (the energy of the hopping integral between the $p_z$ orbitals of two adjacent carbon atoms in a CNT).

## 4. Physisorbed molecules

For completeness we show in figure 4 that the transmission curves close to the Fermi level for CNTs are not affected by the physisorption of molecules, in agreement with it is expected for the weak coupling limit in the tight-binding model.

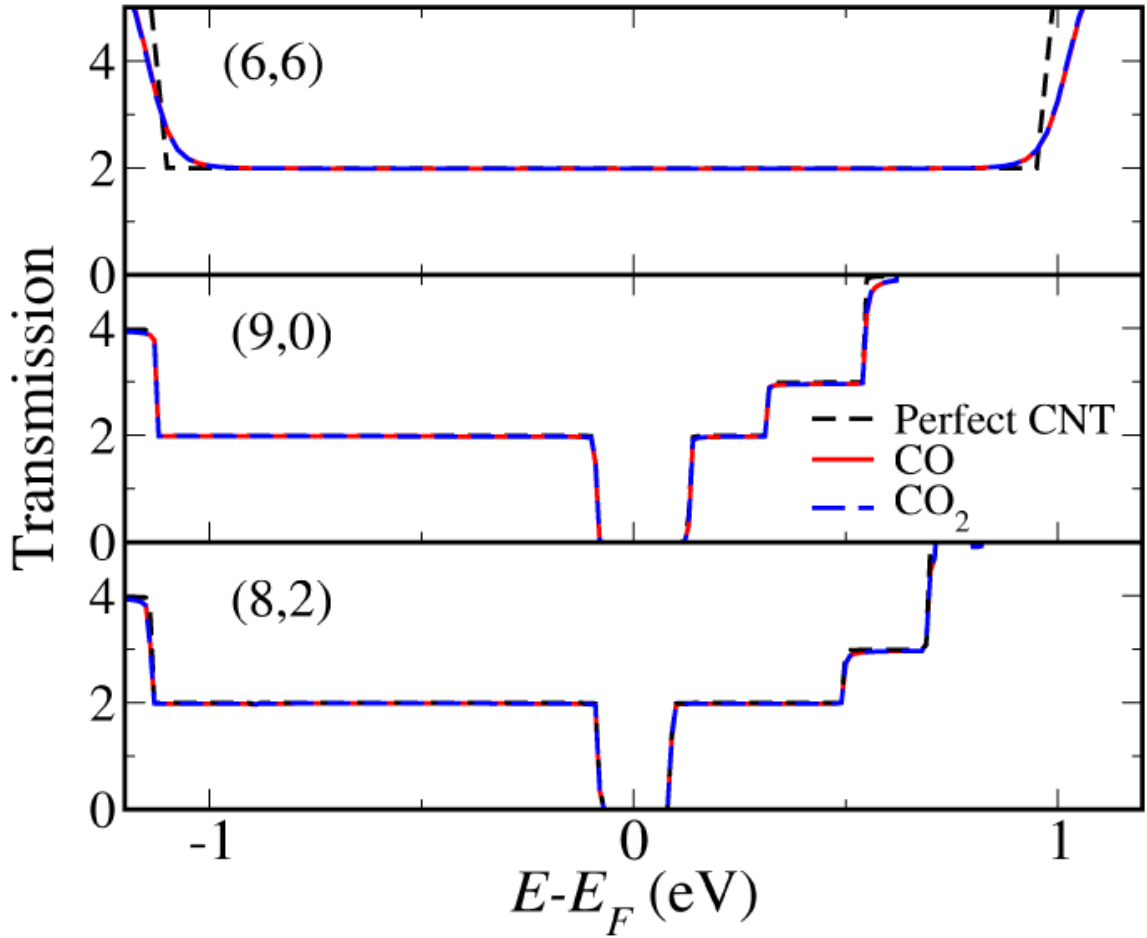

**Figure 4.** Up: Calculated transmission curves (DFT+NEGF) for a (6,6) carbon SWNT doped with one single molecule of CO (solid red) and $CO_2$ (dashed blue), respectively. The transmission curve for the pure CNT is also shown for comparison purposes (dashed black). Central: The same for a (9,0) carbon CNT. Down: The for a (8,2) carbon CNT.

## 5. The case of two adsorbed molecules in the A-B configurations

From the main text it is clear that in this situation we do not get any general trend as the eigenstates of the Hamiltonian are intermixed. Therefore no universal rule can be drawn for the cancellation of the conductance upon molecular absorption. This fact is shown in the figure 5. Lee et al. also discussed four A-B cases in the figure 5 of the ref[14]. In that work (figs 5. a, b, c and d) they treated adsorbed phenyl molecules on top of a carbon atom in a (5,5) CNT. The arrangement of the phenyl pair was similar to our 0-1 and 0-3 cases in our figure 5 (see below) and the trend that they observed for the transmission curves is also coincident with ours.

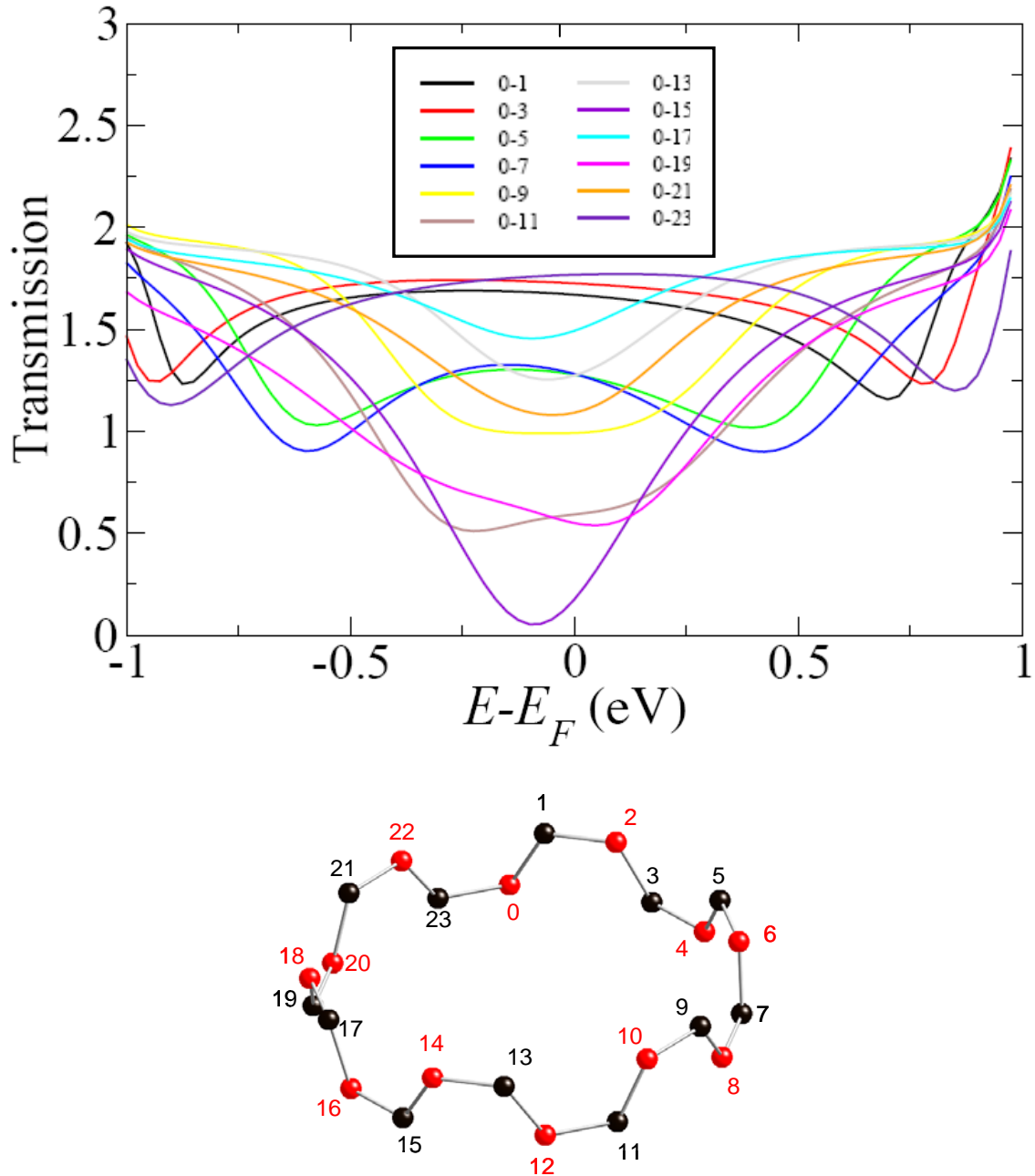

**Figure 5.** Calculated transmission curves (DFT+NEGF) for a (6,6) carbon SWNT doped with two hydrogen atoms in the same unit cell, one of them at an A site (position 0) and the other at a B site (position j). The twelve possible different cases are shown. The lower picture shows the criterion for the labels.

## Acknowledgements

Juan María García-Lastra and Ángel Rubio acknowledge funding Spanish MEC (FIS2007-65702-C02-01), Grupos Consolidados UPV/EHU of the Basque Country Government (IT-319-07) and European Community through e-I3 ETSF project (INFRA-211956); NoE Nanoquanta (NMP4-CT-2004-500198) and SANES (NMP4-CT-2006-017310). Juan María García-Lastra acknowledges funding Spanish MEC through Juan de la Cierva program. Kristian S. Thygesen and Mikkel Strange acknowledge support from The Lundbeck Foundation's Center for Atomic-scale Materials Design and from the Danish Center for Scientific Computing through grant No. HDW-1103-06. We thank Yann Pouillon and Duncan Mowbray for illuminating discussion on this work.